\documentclass[aps,prl,showpacs,twocolumn]{revtex4-1}

\usepackage{graphicx,bbm}

\def\ave#1{\langle #1 \rangle}
\def\dd{{\rm d}}
\begin{document}

\title{Macroscopic Diffusive Transport in a Microscopically Integrable Hamiltonian System}
 
\author{Toma\v z Prosen}
\affiliation{Physics department, Faculty of Mathematics and Physics, University of Ljubljana, Ljubljana, Slovenia}

\author{Bojan \v Zunkovi\v c}
\affiliation{Departamento de F\' isica, Facultad de Ciencias F\' isicas y Matem\' iaticas, Universidad de Chile, Casilla 487-3, Santiago, Chile}

\date{\today}

\begin{abstract}
We demonstrate that a completely integrable classical mechanical model, namely the lattice Landau-Lifshitz classical spin chain, supports diffusive spin transport with a finite diffusion constant in the easy-axis regime, 
while  in the easy-plane regime it displays ballistic transport in the absence of any known relevant local or quasi-local constant of motion in the symmetry sector of the spin current.
This surprising finding should open the way towards analytical computation of diffusion constants for integrable interacting systems and hints on existence of new 
quasi-local classical conservation laws beyond the standard soliton theory.
\end{abstract}

\maketitle

{\em Introduction.-} 
Derivation of irreversible macroscopic transport (e.g. Fourier/Ohm/Fick's) laws from reversible, deterministic, microscopic equations of motion is one of the central questions of statistical physics which  remains largely unsolved even today. It has been believed \cite{ford,casati,livi,bonetto,dhar,gaspard} that chaotic dynamics in classical systems, or more generally strong non-integrability in either quantum or classical systems, are necessary conditions for diffusive transport. Recently, few examples of spin-diffusion at high temperature in completely integrable but strongly interacting quantum spin/particle chains have appeared \cite{gemmer,robin1,spindiff,marko,hubbard}, suggesting that complete integrability might not exclude the possibility of macroscopically diffusive dynamics. It  has remained unclear, however, whether quantum nature of the corresponding many-body dynamics supporting macroscopic entanglement is a necessary condition. Here we show that even quantum correlations are not necessary. By performing extensive numerical simulations in a family of integrable classical spin chains with local interactions -- the {\em lattice Landau-Lifshitz model}~\cite{faddeev} -- we show that spin transport at finite temperature is diffusive in the easy axis regime, while it becomes ballistic in the easy-plane regime and anomalous at the isotropic point. This opens up the possibility for analytic computations of diffusion constants in interacting many-body systems.

Liouville integrability \cite{arnold} is the central concept in the analytic theory of classical mechanics. A Hamiltonian, i.e. conservative system in classical mechanics, is integrable if it possesses the same number of independent conserved quantities as the number of degrees of freedom, call it $n$. In other words, its motion can be reduced to quasi-periodic winding around $n-$dimensional torus embedded in $2n$ dimensional phase space\cite{arnold}. Integrable dynamics is thus regular and manifestly free of sensitive dependence on initial conditions. Nevertheless, integrable systems, though being sparse in nature, represent one of the key topics in mathematical physics as they gave birth to the celebrated soliton theory \cite{faddeev} explaining a variety of observable phenomena, ranging, to name just a few, from localized light in non-linear optics, waves on shallow water and tsunami waves, to elementary particles and localized excitations in condensed matter at low temperatures.
 
The solitons, indestructible localized packages of energy which propagate through the system and scatter from each other like elastic hard balls, have been believed to be the reason why integrable extended systems behave as ideal - ballistic conductors of heat, particles, electric charge, magnetization, etc. \cite{ford,casati}. Being particularly interested in the one-dimensional lattice systems, where $n$ particles are arranged along a line or a ring such that only nearest neighbors can interact representing the simplest model of crystalline solids, one finds that existence of non-trivial conservation laws (besides the transported quantity, e.g. energy, particle number, electric charge, magnetization) generically implies the ballistic (non-diffusive) transport \cite{zotos}. This statement, which builds on an old idea of Mazur \cite{mazur}  but has only recently been formally proven \cite{ilievski}, essentially states the following. Whenever there exist a quantity $I$ which is conserved
in time $I(t)\equiv I$ and independent of the transported quantity itself, such that $\ave{I J} \neq 0$, where $J(t)=\sum_{x=1}^n j(x,t)$ is the current with $j(x,t)$ being the current density at time $t$ and at site $x$ in the lattice and $\ave{\bullet}$ denotes the thermodynamic average (for fixed, specified values of temperature, electro-chemical potential,  magnetization, etc.), then the transport is ballistic and the corresponding Kubo conductivity $\kappa$ diverges. Conductivity is related to a diffusion constant $D$, via the generalized Einstein relation $\kappa = D/T$ where $T$ is the absolute temperature, and the latter can be within Green-Kubo linear response theory expressed
\begin{equation}
D = \lim_{\tau\to\infty} \lim_{n\to\infty} \sum_{r=1}^n \int_0^\tau C(r,t) {\rm d}t,
\end{equation}
in terms of the integrated spatio-temporal current-current correlation function  $C(r,t) = \ave{j(x,0) j(x+r,t)}$.
Lattice site $x$ is arbitrary for translationally invariant systems which exhaust major examples of integrable systems. Note that in statistical mechanics the thermodynamic limit of size $n\to\infty$ has to be taken always before the time $\tau\to\infty$ limit. It is clear that the above expression for the diffusion constant can be given in terms of time-correlations only, namely
$D = \lim_{t \to \infty} D(t)$, introducing a time-dependent diffusion constant as
\begin{equation}
D(\tau)= \int_0^\tau C(t) {\rm d}t
\label{eq:Dt}
\end{equation} 
where we can write the total current time autocorrelation as $C(t) = \frac{1}{n}\ave{J(0) J(t)}$. The existence of solitons and nontrivial conserved quantities in integrable systems implies non-vanishing tails of the time correlations\cite{zotos,ilievski} $C_\infty = C(t\to\infty) \neq 0$, in turn  implying linear divergence of the time-dependent diffusion constant $D \to C_\infty t \to \infty$ which is a signature of ballistic transport. Up to date, all studies of transport in classical integrable particle chains have persistently showed ballistic transport (see e.g. Ref.~\cite{david} and references therein).
On the other hand, for diffusive transport ($D < \infty$), we have evidence that not even microscopic chaos \cite{gaspard} is neccessary but a weaker property of  dynamical mixing \cite{sinai} is sufficient \cite{dettmann,david,baowen}. Furthermore, some recent numerical studies of quantum spin $1/2$ chains with anisotropic Heisenberg interaction ($XXZ$ chains) indicated \cite{robin1,spindiff,robin2,fabian} that the high temperature spin transport is diffusive in the easy-axis regime, despite the fact that the $XXZ$ chains are quantum integrable by the algebraic Bethe ansatz which is the quantum version of the soliton theory. These results have been further corroborated with evidence of particle and spin diffusion in another Bethe Ansatz integrable model namely the one-dimensional Hubbard model \cite{hubbard}. Nevertheless, as the completeness of Bethe ansatz solutions has not been proven in these models and as quantum dynamics supports a high degree of complexity as opposed to classical dynamics due to exponentially large (in $n$) Hilbert space dimension and possibility of macroscopic entanglement \cite{vedral}, it has remained unclear whether the key to the observed diffusion really lies in the integrability structures of the Heisenberg and Hubbard models.

\begin{figure}[htb]
\begin{center}
\vspace{2mm}
	\includegraphics[width=0.48\textwidth]{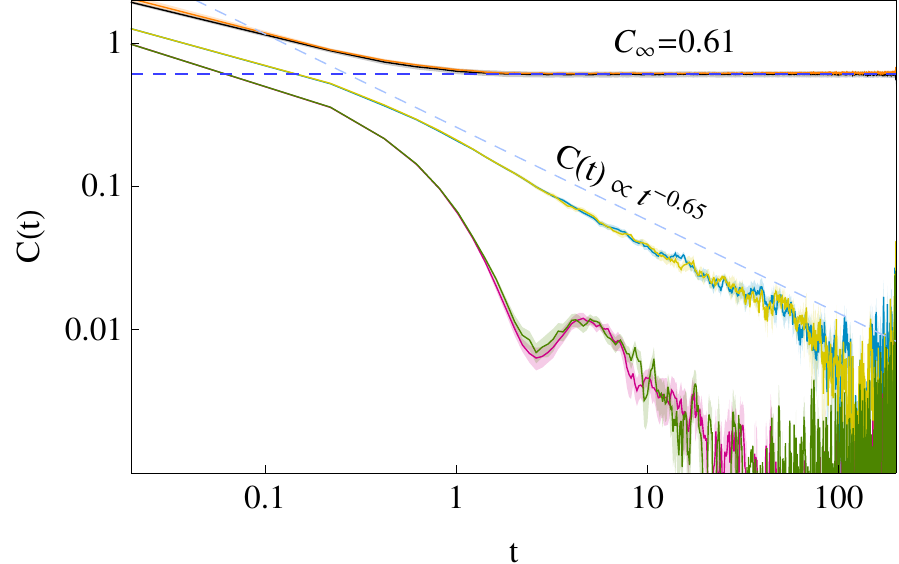}
\caption{(Color online) Current autocorrelation function $C(t)$ in log-log scale for easy-plane regime (top curves, orange: $n=160$, black: $n=2560$), isotropic regime (middle curves, yellow: $n=2560$, blue: $n=5120$) and easy-axis regime (bottom curves, violet: $n=2560$, green: $n=5120$). Shaded regions denote the estimated statistical error for ensemble averages over $N\approx 10^3$ initial conditions. Dashed lines denote asymptotic behavior for large time in the easy-plane regime (dark-blue) and isotropic regime (light-blue).}
\label{Cfig}
\end{center}
\end{figure}
\begin{figure}[htb]
\begin{center}
	\includegraphics[width=0.2495\textwidth]{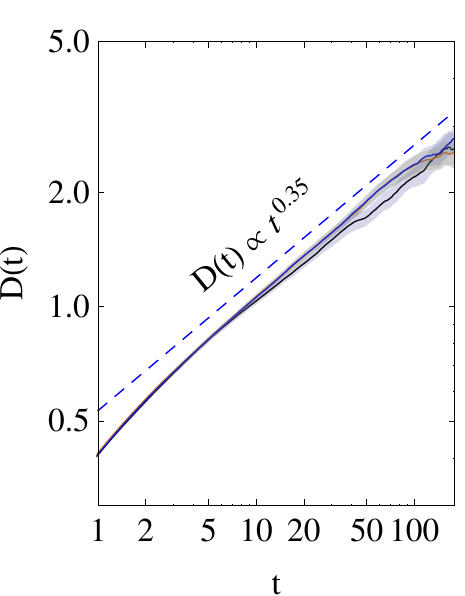}$\!\!\!\!\!\!$ \includegraphics[width=0.2455\textwidth]{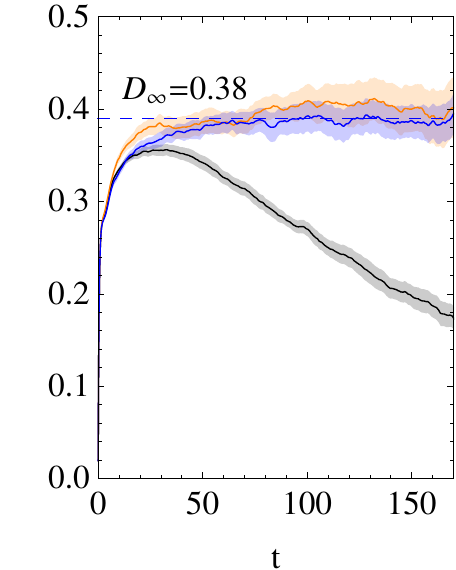}
\caption{(Color online) Time dependent diffusion constant $D(t)$ in isotropic regime (left) and easy-axis regime (right) comparing different system sizes $n=640$ (black), $n=2560$ (orange) and $n=5120$ (blue). 
The shaded region denotes estimated statistical error and the dashed lines show estimated asymptotic behavior. }
\label{Dfig}
\end{center}
\end{figure}

 {\em Model and methods.-}
However, in this Letter we show a convincing numerical evidence for spin diffusion at high temperature in a fully classical completely integrable model. We consider  a lattice Landau-Lifshitz (LLL) model \cite{faddeev} for a chain or a ring of $n$ classical spins described by angular-momentum vectors $\vec{S}_x$, $x=1,\ldots n$ of equal fixed length $|\vec{S}_x| \equiv R$ and the Hamiltonian function
of the form 
\begin{equation}
H=\sum_{x=1}^n h(\vec{S}_x,\vec{S}_{x+1}),
\label{eq:Ham}
\end{equation}
where the nearest-neighbor spin interaction density $h(\vec{S},\vec{S}')$ is given by
\begin{widetext}
\begin{equation}
h(\vec{S},\vec{S}') = \log\left|\cosh(\rho S_3)\cosh(\rho S'_3)+ \coth^2(\rho R)\sinh(\rho S_3)\sinh(\rho S'_3) + \sinh^{-2}(\rho R)F(S_3)F(S'_3)(S_1 S'_1 + S_2 S'_2)\right|
\label{eq:hh}
\end{equation}
and $F(S)\equiv \sqrt{(\sinh^2 (\rho R) - \sinh^2(\rho S))/(R^2 - S^2)}$ (as defined by Faddeev and Takhtajan, Ref.\cite{faddeev}, Chapter III.5).
$\rho$ is the model's parameter, which can be real or purely imaginary. So we shall reparametrize it with an alternative {\em real} parameter $\delta = \rho^2$, which may -- in analogy with the closely related $XXZ$ spin chain \cite{doikou} -- be called {\em anisotropy parameter}. The cases with $\delta > 0$ correspond to {\em easy-axis} spin-spin interaction, those with $\delta < 0$ to {\em easy-plane} interaction, whereas $\delta=0$ designates the case of {\em isotropic} interaction. 

As the third component $M_3$ of the total magnetization  $\vec{M} = \sum_{x=1}^n \vec{S}_x$ is a constant of motion, one finds that 
 the equation of motion for local magnetization ${\rm d} S_{x,3}/{\rm d} t = j_{x} - j_{x-1}$ has a form of a local continuity equation from which we read out the expression for the spin current density $j_x = j(\vec{S}_x,\vec{S}_{x+1})$ in LLL model, namely
\begin{equation}
j(\vec{S},\vec{S}') = \sinh^{-2}(\rho R) (S_2 S_1' - S_1 S_2')F(S_3)F(S'_3)\exp\bigl[-h(\vec{S},\vec{S}')\bigr].
\label{eq:jj}
\end{equation}
\end{widetext}
The Hamiltonian and the current density in the easy-plane regime $\delta < 0$ can be obtained from (\ref{eq:hh},\ref{eq:jj}) by analytic continuation, i.e., by replacing $\cosh(\rho \bullet)$ by $\cos(s \bullet)$ and $\sinh(\rho \bullet)$ by 
$\sin (s \bullet)$ where $s=\sqrt{-\delta}$. 
In the isotropic case $\delta=0$ one performs the limit $\rho \to 0$ in (\ref{eq:hh},\ref{eq:jj}) and obtains explicitly 
\begin{equation}
h(\vec{S},\vec{S}') = \log \left( 1 + \frac{\vec{S}\cdot \vec{S}'}{R^2}\right),\; j(\vec{S},\vec{S}') = \frac{S_2 S_1' - S_1 S_2'}{R^2 + \vec{S}\cdot \vec{S}'}. \label{eq:hi}
\end{equation}
Time development of each spin is given by solving Hamilton's equations ${\rm d}\vec{S}_x/{\rm d}t = \frac{\partial H}{\partial \vec{S}_{x}} \times \vec{S}_x$, 
where $\times$ denotes the cross product.
These equations of motion for $\vec{S}_x(t)$ are then solved using a variable step integrator of {\sc Matlab} with required relative accuracy of $10^{-6}$ for all trajectories. Besides checking the accuracy of trivial conservation laws, such as $H$ and $M_3$, we have also checked that numerically determined Lyapunov exponents \cite{arnold} vanish asymptotically in all three regimes ($\delta >, =, < 0$) as required for a completely integrable system.
 
We chose $N$ initial conditions generated by the Metropolis algorithm \cite{Mnote} yielding a thermal Gibbs ensemble with given temperature $T$  and vanishing $M_3=0$ component (along the symmetry axis of the spin-spin interaction) of the magnetization. We note that ensembles with non-vanishing or non-fixed $M_3$ would not yield the correct Kubo formula \cite{bonetto} for the spin-diffusion constant in the absence of external magnetic field. The LLL model is invariant under a canonical transformation 
\begin{equation}
 {\cal R} : (S_{x,1},S_{x,2},S_{x,3})\to (S_{x,1},-S_{x,2},-S_{x,3}),
 \end{equation} 
 i.e. a $\pi$-rotation around the first cartesian axis, namely ${\cal R}H = H$. As the LLL $r-$matrix \cite{faddeev} is invariant under ${\cal R}$ so are also all the derived local conserved quantities ${\cal R}I_k = I_k$. However, the spin current is {\em odd} under the symmetry transformation, ${\cal R}J = -J$, so all (ballistic) terms  in the  Mazur bound \cite{zotos} have to vanish, $\ave{I_k J} = \ave{{\cal R}I_k {\cal R}J} = -\ave{I_k J}$, allowing for the possibility of non-ballistic transport.

\begin{widetext}

\begin{figure}[!htb]
\begin{center}
	\hspace{-0.9cm}\includegraphics[width=0.32\textwidth]{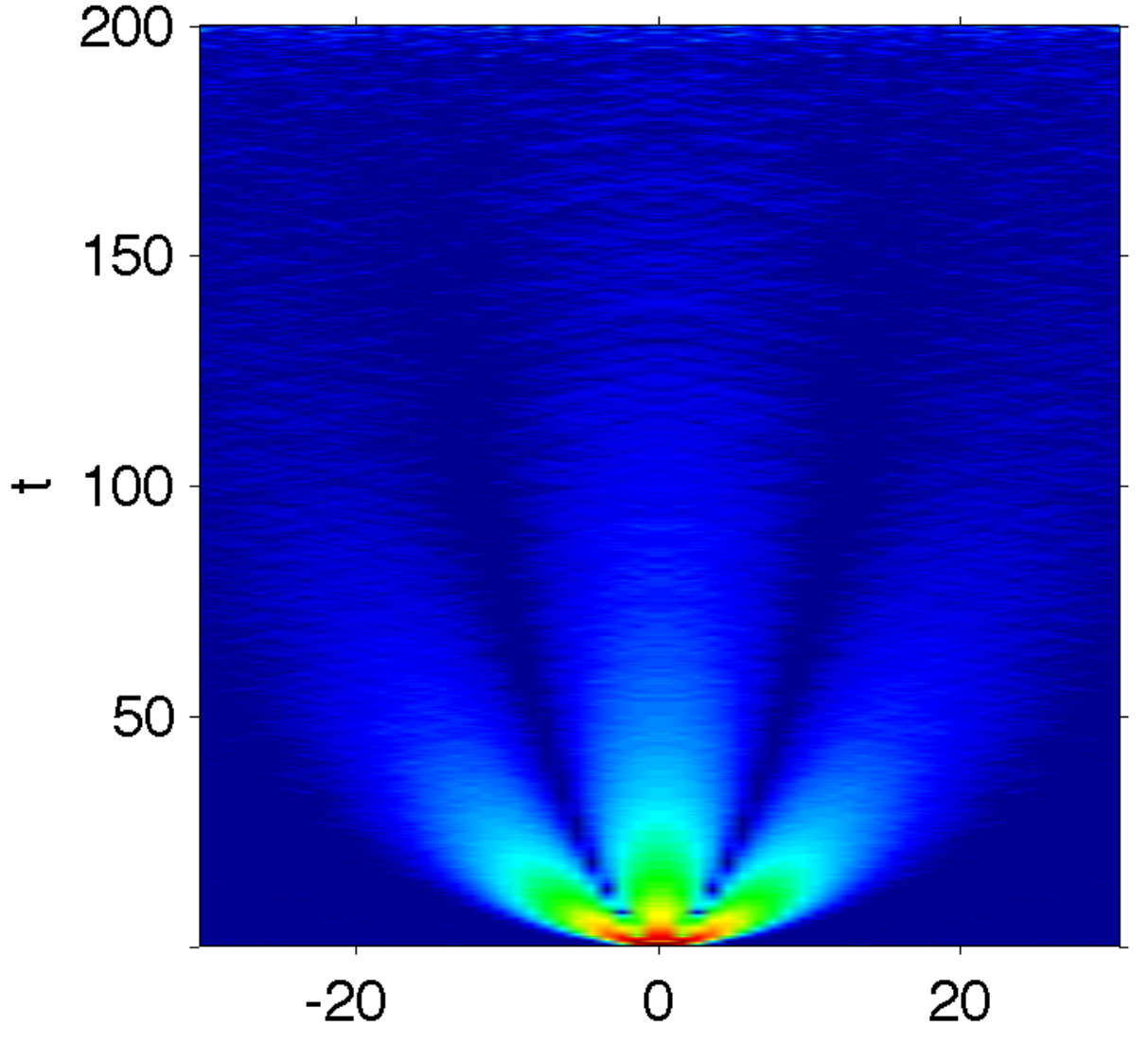}~~\includegraphics[width=0.303\textwidth]{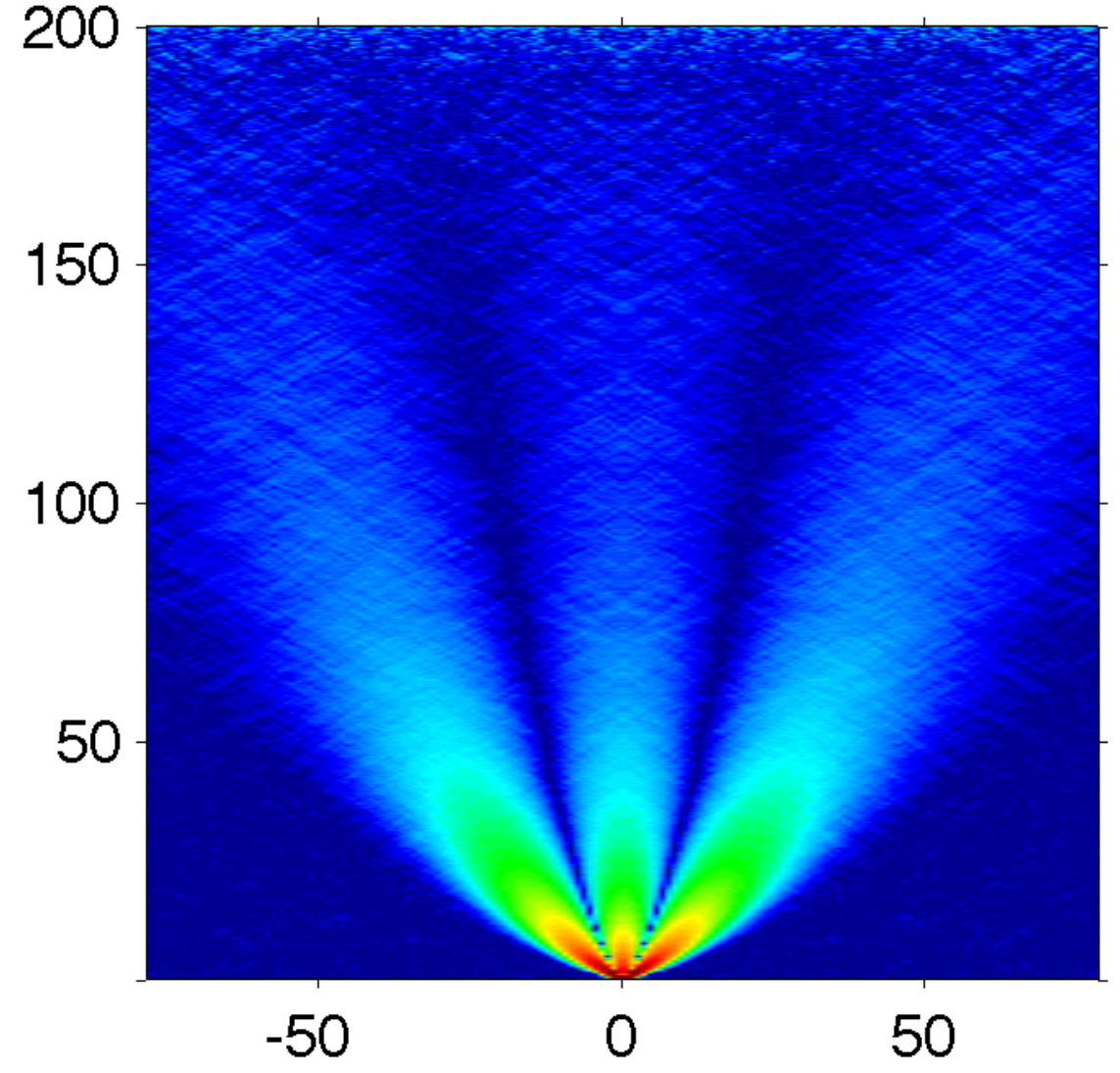}~~\includegraphics[width=0.300\textwidth]{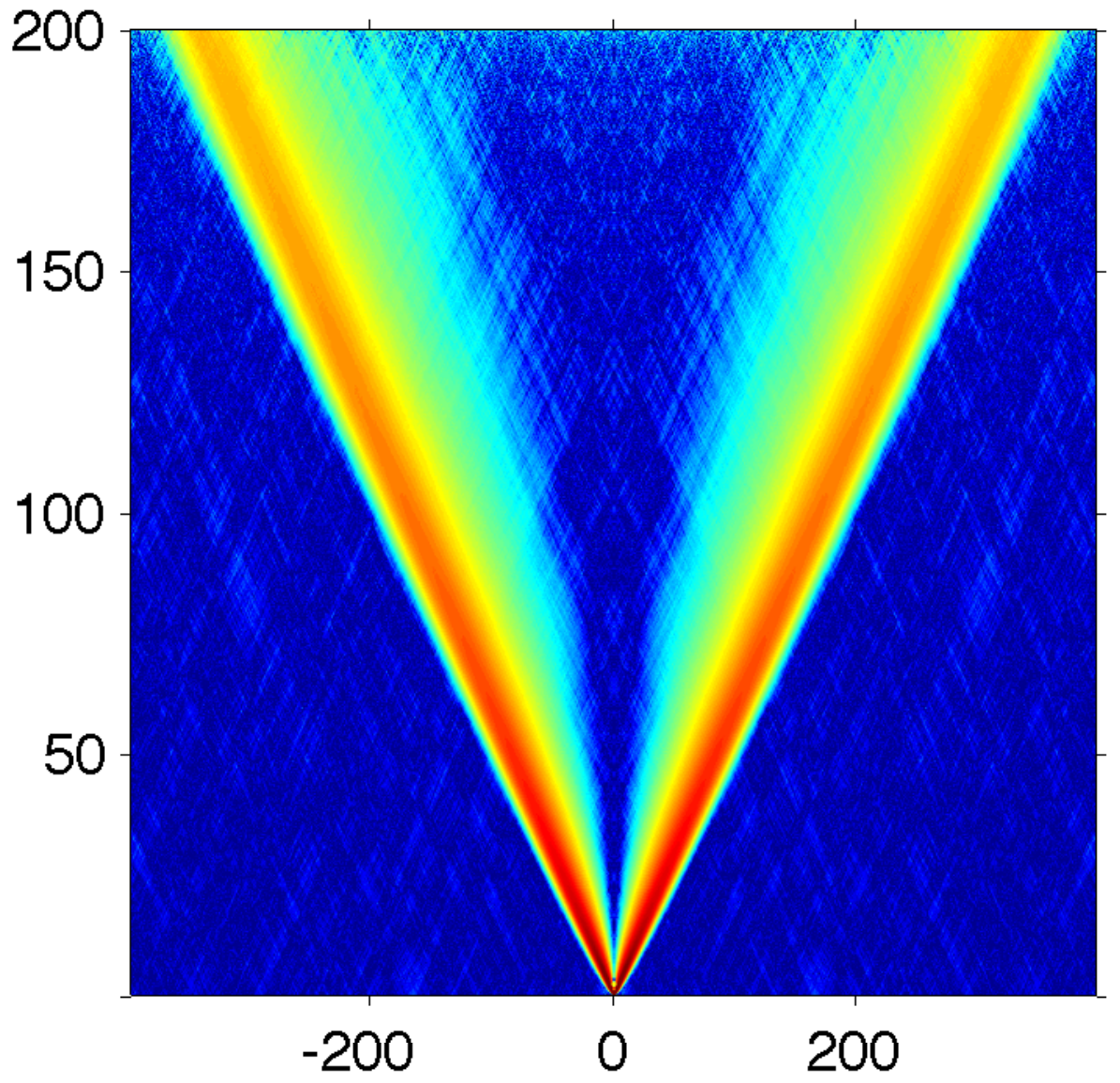}
	
	\includegraphics[width=1\textwidth]{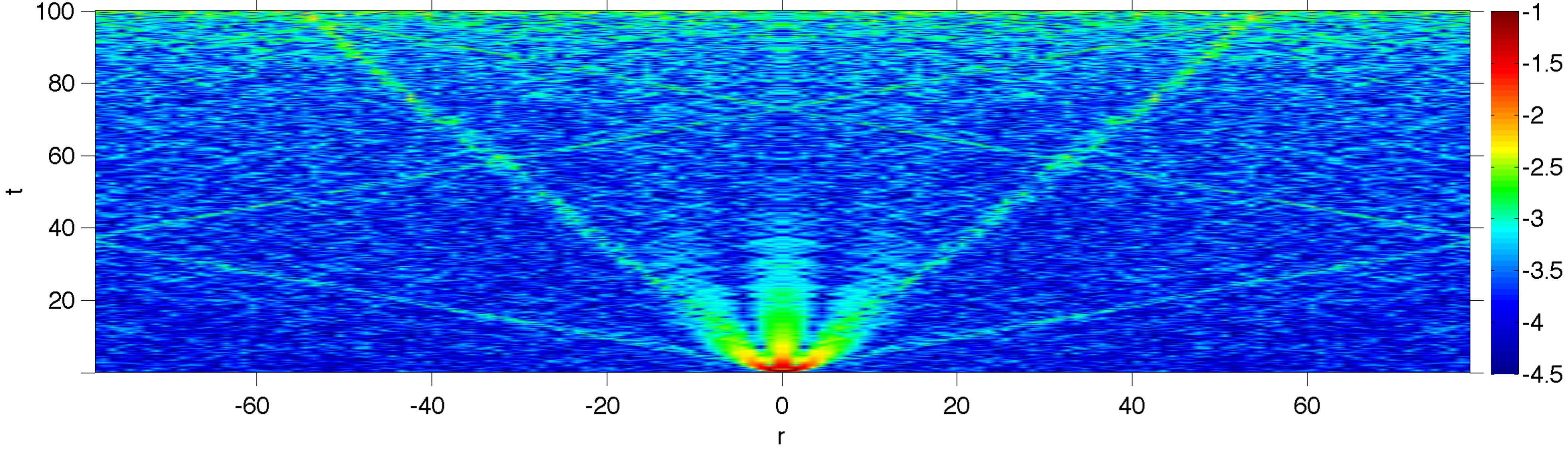}
\caption{(Color online) Modulus of spatiotemporal spin current autocorrelation function $|C(r,t)|$ shown in log-scale with color scale ranging from $10^{-4.5}$ to $10^{-1}$ indicated in the bottom-right. In the upper panels we show data averaged over ensembles of $N\approx 10^3$ initial conditions in easy-axis (left; $n=5120$), isotropic (center; $n=5120$ ) and easy-plane (right; $n=2560$) regimes. The bottom plot shows easy-axis regime again with a smaller system/ensemble size $n=160, N=600$  where {\em scars} of solitons emerging from local thermal fluctuations are still clearly visible.}
\label{STfig}
\end{center}
\end{figure}

\end{widetext}

{\em Results.-}
Solving Hamilton's equations numerically we compute the spatio-temporal current-current autocorrelation function $C(r,t)$, as well as  the temporal autocorrelation $C(t)$ and obtain the time-dependent spin diffusion constant $D(t)$ (\ref{eq:Dt}). We chose several different chain sizes $n$ (up to $n=5120$) and ensemble size $N$ large enough (typically $N\sim 10^3$) so that finite size effects, and the statistical error scaling as $\sim 1/\sqrt{nN}$, appear negligible. Representative values of the anisotropy parameters are chosen as $\delta=1$ (easy-axis), $\delta=-1$ (easy-plane), and $\delta=0$ (isotropic regime) while fixing $R=1$. In all cases the temperature of the initial Gibbs ensemble is $T=4$. In Fig.~\ref{Cfig} we plot spin current autocorrelation $C(t)$ in all three regimes and find a finite saturation value $C_\infty \neq 0$ implying ballistic transport in the easy-plane regime, whereas, in the isotropic and easy-axis regimes the tails of the current autocorrelation function are vanishing. To elaborate on the tails of the current autocorrelation function we plot  in Fig.~\ref{Dfig} the time dependent diffusion constant $D(t)$, which in the isotropic regime shows power low behavior $D(t)\propto t^\alpha$ with $\alpha\approx0.35$ and in the easy-axis regime saturates at a finite value $D(\infty)\approx0.38$, providing a firm evidence of anomalous spin-transport in the isotropic regime and diffusive spin-transport in the easy-axis regime. These results can be even better illustrated by portraying spatio-temporal correlations $C(r,t)$ in Fig.~\ref{STfig}. In the easy-plane regime we find a clear causality-cone structure with non-decaying tails whereas in the easy-axis case the cone is curved inwards and the tails are strongly damped in time resulting in suppression of all ballistic contributions for long times.  In bottom panel of Fig.~\ref{STfig} we present data for much smaller ensemble of initial conditions and smaller system size showing ``scars'', i.e. ballistic {\em solitons} which may emerge in initial conditions with localized large thermal fluctuations and which can travel much faster than the correlation velocity, but contributions of which (due to the existence of ${\cal R}$ symmetry) vanish in the limit of infinitely large ensemble $N\to\infty$. Therefore, one would recognize the fact that we are dealing with a completely integrable system (in particular in the easy axis regime) only by looking at a {\em finite} $N$ data.

{\em Conclusion.-} The surprising possibility of having a regime with normal, diffusive transport in completely integrable classical system (as found in the easy-axis regime of LLL model), which is associated with the existence of parity-type symmetry of the classical r-matrix whose operation changes the sign of the corresponding transporting current, opens up the immediate question if the classical r-matrix theory \cite{faddeev} can be updated to allow for analytical calculations of diffusive transport coefficients. Equally interesting is the problem of explaining the ballistic regime in such a situation (say in the easy-plane regime of LLL model), since the Mazur bound identically vanishes. As LLL model can be considered as a classical limit of $XXZ$ spin chain (see also \cite{doikou}), we conjecture the existence of a classical analog of the construction of quasi-local conserved quantities with negative parity \cite{prosen}  going beyond the standard soliton theory \cite{faddeev}.

\section*{Acknowledgments}

TP warmly thanks Pavle Saksida for stimulating discussions on classical integrable systems.
The work has been supported by the grant P1-0044 of the Slovenian Research Agency (TP) and Fondecyt project 3130495 (B\v Z).


\begin{thebibliography}{1}

\bibitem{ford}  J.~Ford, {\em The Fermi-Pasta-Ulam Problem: Paradox turns Discovery}, Phys. Rep. {\bf 213}, 271-310 (1992).

\bibitem{casati} G.~Casati, J.~Ford, F.~Vivaldi, and W.~H.~Visscher, {\em One-dimensional classical many-body
system having normal thermal conductivity}, Phys. Rev. Lett. {\bf 52}, 1861-1864 (1984).

\bibitem{livi} R.~Livi, A.~Politi and S.~Lepri, {\em Thermal conduction in classical low-dimensional lattices}, Phys. Rep. {\bf 377}, 1-80 (2003).

\bibitem{bonetto} F.~Bonetto, J.~L.~Lebowitz and L.~Rey-Bellet, {\em Fourier law: a challenge to theorists}, in Mathematical
Physics 2000, A. Fokas, A. Grigoryan, T. Kibble, and B. Zegarlinski, eds. (Imperial College Press, London, 2000), pp. 128150.

\bibitem{dhar} A.~Dhar, {\em Heat Transport in low-dimensional systems}, Adv. Phys., {\bf 57}, 457-537 (2008).

\bibitem{gaspard} P.~Gaspard, M.~E.~Briggs, M.~K.~Francis, J.~V.~Sengers, R~W.~Gammon, J.~R.~Dorfman abd R.~V.~Calabrese, {\em Experimental evidence for microscopic chaos}, Nature {\bf 394}, 865-868 (1998).

\bibitem{gemmer} R.~Steinigeweg and J.~Gemmer, {\em  Density dynamics in translationally invariant spin-1/2 chains at high temperatures: A current-autocorrelation approach to finite time and length scales}, Phys. Rev. B. {\bf 80}, 184402 (2009).

\bibitem{robin1} R.~Steinigeweg, {\em Decay of currents for strong interactions}, Phys. Rev. E {\bf 84}, 011136 (2011).

\bibitem{spindiff} T.~Prosen and M. \v Znidari\v c, {\em Matrix product simulation of non-equilibrium steady states of quantum spin chains}, J. Stat. Mech.: Theory Exp., {\bf 2009}, P02035 (2009).

\bibitem{marko} M.~\v Znidari\v c, {\em Spin transport in a one-dimensional anisotropic Heisenberg model}, Phys. Rev. Lett. {\bf 106}, 220601 (2011).

\bibitem{hubbard} T. Prosen and M. \v Znidari\v c, {\em Diffusive high-temperature transport in the one-dimensional Hubbard model}, Phys. Rev. B {\bf 86}, 125118 (2012).

\bibitem{faddeev} L.~D.~Faddeev and L.~A.~Takhtajan, {\em Hamiltonian Methods in the Theory of Solitons}, (Springer-Verlag, 1987).

\bibitem{arnold} V.~I.~Arnold, {\em Mathematical Methods of Classical Mechanics}, (Springer-Verlag, 1989).

\bibitem{zotos} X.~Zotos, F.~Naef and P.~Prelov\v sek, {\em Transport and conservation laws}, Phys. Rev. B {\bf 55}, 11029 (1997).

\bibitem{mazur} P. Mazur, {\em Non-ergodicity of phase functions in certain systems}, Physica {\bf 43}, 533-545 (1969).

\bibitem{ilievski} E.~Ilievski and T. Prosen, {\em Thermodynamic bounds on Drude weights in terms of almost-conserved quantities}, Commun. Math. Phys. {\bf 318}, 819 (2013).

\bibitem{david}  T.~Prosen and D.~K.~Campbell, {\em Normal and anomalous heat transport in one-dimensional classical lattices}, Chaos {\bf 15}, 015117 (2005).

\bibitem{sinai} I.~P.~Cornfeld, S.~V.~Fomin and Ya.~G.~Sinai, {\em Ergodic theory}, (Springer-Verlag, 1982).

\bibitem{dettmann} C.~P.~Dettmann, E.~G.~D.~Cohen and H.~van Beijeren, {\em Microscopic chaos from Brownian motion?}, Nature {\bf 401}, 875 (1999).

\bibitem{baowen} B.~Li, G. Casati, J.~Wang and T.~Prosen, {\em  Fourier Law in the Alternate-Mass Hard-Core Potential Chain}, Phys. Rev. Lett. {\bf 92}, 254301 (2004).

\bibitem{robin2} R.~Steinigeweg, {\em Spin transport in the XXZ model at high temperatures: Classical dynamics versus quantum S=1/2 autocorrelations}, Europhys. Lett. {\bf 97}, 67001 (2012).

\bibitem{fabian} S.~Langer, F.~Heidrich-Meisner, J.~Gemmer, I.~P.~McCulloch, and U.~Schollw\" ock, {\em Real-time study of diffusive and ballistic transport in spin-1/2 chains using the adaptive time-dependent density matrix renormalization group method}, Phys. Rev. B {\bf 34}, 214409 (2009).

\bibitem{vedral} L.~Amico, R.~Fazio, A.~Osterloh, V.~Vedral, {\em Entanglement in many-body systems}, Rev. Mod. Phys. {\bf 80}, 517-576 (2008).

\bibitem{doikou} J.~Avan, A.~Doikou and K.~Sfetsos, {\em Integrable quantum spin chains and their classical continuous counterparts}, Nucl. Phys. B {\bf 840}, 469 (2010).

\bibitem{Mnote} The ensemble of initial conditions is sampled via a standard Metropolis algorithm \cite{metropolis}, where the new configuration in each Metropolis step is generated by choosing two random spins  and change their values randomly in the $\phi$-$z$ plane ($\phi$ denotes the angle of a spin-vector in the $xy$ plane and $z$ is its component along the $z$-axis) according to a uniform probability with a constraint  preserving the total magnetization (assuring $M_3=0$), namely $\dd P \propto \dd \phi_1 \dd \phi_2 \dd z_1 \dd z_2 \delta(z_1+z_2-(z'_1+z'_2))$, where $z'_{1,2}$ denote the spin components before the change. 
The acceptance probability of the step is given by the Gibbs factor $\exp((E'-E)/ T)$, where $E'$ and $E$ denote the values of $H$, Eqs. (\ref{eq:Ham},\ref{eq:hh},\ref{eq:hi}), of the configurations before and after the step, respectively, and $T$ is the temperature.

\bibitem{metropolis} N.~Metropolis, A.~W.~Rosenbluth, M.~N.~Rosenbluth, A.~H.~Teller, E.~Teller, {\em Equations of State Calculations by Fast Computing Machines}, J. Chem. Phys. {\bf 21}, 1087Ð1092 (1953).

\bibitem{prosen} T.~Prosen, {\em Open XXZ Spin Chain: Nonequilibrium Steady State and a Strict Bound on Ballistic Transport}, Phys. Rev. Lett. {\bf 106}, 217206 (2011).

\end{thebibliography}
\end{document}